\documentclass[prd,showpacs,showkeys,preprintnumbers,floatfix,groupedaddress,
nofootinbib,superscriptaddress,twocolumn]{revtex4-1}
\usepackage{hyperref} 
\usepackage{graphicx}
\usepackage{amssymb}
\usepackage{amsmath}
\usepackage{bbold}
\usepackage{epstopdf}
\usepackage{float}
\usepackage{caption}
\usepackage{slashed}
\usepackage{verbatim}
\usepackage[singlelinecheck=false]{subcaption}

\DeclareGraphicsExtensions{.pdf,.png,.jpg}
\DeclareMathOperator{\tr}{tr}
\def\chpt{\raise0.4ex\hbox{$\chi$}PT}
\def\mhat{\widehat m}
\def\mqhat{\widehat m_q}

\def\ephat{\widehat \epsilon}
\def\epqhat{\widehat \epsilon_q}

\def\alpEM{\alpha_{\rm EM}}

\def\LOP{LO${}^+$}
\def\EM{{\rm EM}}
\def\cEM{c_{\rm EM}}
\def\QCD{{\rm QCD}}
\def\thetaQCD{\theta_{\rm QCD}}
\def\PQ{{PQ}}

\begin{document}

\title{Impact of electromagnetism on
phase structure for Wilson and twisted-mass fermions
including isospin breaking}

\author{Derek P. Horkel}
\email[e-mail: ]{dhorkel@uw.edu}
\affiliation{
 Physics Department, University of Washington, 
 Seattle, WA 98195-1560, USA \\
}
\author{Stephen R. Sharpe}
\email[e-mail: ]{srsharpe@uw.edu}
\affiliation{
 Physics Department, University of Washington, 
 Seattle, WA 98195-1560, USA \\
}
\date{\today}
\begin{abstract}
In a recent paper we used chiral perturbation theory to determine
the phase diagram and pion spectrum for
Wilson and twisted-mass fermions at nonzero lattice spacing
with nondegenerate up and down quarks.
Here we extend this work to include the effects of electromagnetism,
so that it is applicable to recent simulations incorporating
all sources of isospin breaking.
For Wilson fermions, we find that the phase diagram is unaffected
by the inclusion of electromagnetism---the only effect is to
raise the charged pion masses.
For maximally twisted fermions, we previously took the
twist and isospin-breaking directions to be different, in order that
the fermion determinant is real and positive. However, 
this is incompatible with electromagnetic gauge invariance, 
and so here we take
the twist to be in the isospin-breaking direction,
following the RM123 collaboration.
We map out the phase diagram in this case, which has not 
previously been studied.
The results differ from those obtained with different twist and isospin
directions. 
One practical issue when including electromagnetism is that the critical
masses for up and down quarks differ. We show that one of the criteria
suggested to determine these critical masses does not work, and propose
an alternative.
\end{abstract}

\maketitle

\section{Introduction}
\label{sec:intro}

The phase diagram of lattice QCD (LQCD) 
can contain unphysical transitions and unwanted phases
due to discretization effects.
A well known example is the Aoki phase that can be present
with Wilson-like fermions~\cite{Aoki:1983qi}.\footnote{%
``Wilson-like'' refers to both unimproved and improved versions
of Wilson fermions. The choice will not matter in this work.}
Unphysical phases occur when the effects of
physical light quark masses are comparable to those induced
by discretization,
specifically $m \sim a^2\Lambda_{\rm QCD}^3$, with $a$ the lattice spacing.
This can be shown by extending chiral perturbation theory (\chpt)
to include the effects of discretization~\cite{Sharpe:1998xm}.
Understanding the phase structure is necessary so that
LQCD simulations can avoid working close to
unphysical phases, so as to avoid distortion of results and
critical slowing down.

Recently, we extended the analysis of the 
phase diagram to the case of nondegenerate up and
down quarks for Wilson-like and twisted-mass fermions~\cite{Horkel:2014nha}.
This was prompted by the recent incorporation of mass splittings
into simulations of LQCD.\footnote{%
For recent reviews of such simulations
see Refs.~\cite{Portelli:2013jla,Tantalo:2013maa}.}
We found a fairly complicated phase structure, in which,
for example, the
Aoki phase was continuously connected to Dashen's CP-violating
phase~\cite{Dashen:1970et,Creutz:2003xu}.

A drawback of our analysis was that it did not include
the other major source of isospin breaking in QCD, namely
electromagnetism.
For most hadron properties, electromagnetic effects are comparable
to those of the mass nondegeneracy $\epsilon_q=(m_u\!-\!m_d)/2$.
For example, in the neutron-proton mass difference these
two effects lead to contributions of approximately
$-1\;$MeV and $2.5\;$MeV, respectively.\footnote{%
These results are from the recent LQCD calculation of 
Ref.~\cite{Borsanyi:2014jba}, and use the convention of that work
for the separation of electromagnetic and $\epsilon_q$ effects.}
Furthermore, the recent LQCD simulations alluded to above
have included both mass nondegeneracy and electromagnetism.
Thus, to be directly applicable to such simulations, we must
extend our analysis to include electromagnetism.
This is the purpose of the present note.

We work in Wilson or twisted-mass \chpt\
(both of which we refer to as W\chpt\ for the sake of brevity)
using a power-counting to be explained in Sec.~\ref{sec:power}.
At the order we work, 
it turns out that the inclusion of electromagnetism can be
accomplished in most cases simply by shifting low-energy coefficients (LECs)
in the results without electromagnetism.
Thus we can take over many results from Ref.~\cite{Horkel:2014nha}
without further work.

One new issue concerns the simultaneous 
inclusion of electromagnetism and
quark nondegeneracy with twisted-mass fermions.
The approach we used in the absence of electromagnetism
in Ref.~\cite{Horkel:2014nha}
(following Ref.~\cite{Frezzotti:2003xj}) was to
apply the twist in a different direction in isospin space ($\tau_1$)
from that in which the masses are split ($\tau_3$).
This leads to a real quark determinant, and is the method used
to simulate the $s$ and $c$ quarks using twisted-mass fermions
(see, e.g., Ref.~\cite{Carrasco:2014cwa}).
This does not, however, generalize to include electromagnetism 
in a gauge-invariant way.
Here, instead, we follow Ref.~\cite{deDivitiis:2013xla}, 
and twist in the $\tau_3$ direction.
When doing simulations, this has the disadvantage 
of leading to a complex quark determinant,\footnote{%
This is avoided in 
Refs.~\cite{deDivitiis:2011eh,deDivitiis:2013xla} by expanding
about the theory with degenerate quarks and no electromagnetism.}
but there are no barriers to studying the theory with \chpt.

The remainder of this paper is organized as follows.
We begin in Sec.~\ref{sec:power} with
a brief discussion of our power-counting scheme and
a summary of relevant results from Ref.~\cite{Horkel:2014nha}.
We then explain, in Sec.~\ref{sec:Wilson},
how electromagnetism changes the results of Ref.~\cite{Horkel:2014nha}
for the case of Wilson-like fermions.
Section~\ref{sec:twistEM} describes how to 
simultaneously include isospin breaking, electromagnetism and twist,
while Sec.~\ref{sec:twistchiral} gives our corresponding results for the
phase diagram, focusing mainly on the case of maximal twist.
We conclude in Sec.~\ref{sec:conc}.

Two technical issues are discussed in appendices.
The first concerns the renormalization factors needed to
relate lattice masses to the continuum masses that
appear in \chpt. This issue is subtle because
singlet and nonsinglet masses renormalize differently.
This point was not discussed in Ref.~\cite{Horkel:2014nha},
and we address it in Appendix~\ref{app:renorm},
except that we do not include all the effects introduced by
electromagnetism.

The second appendix concerns the need for charge-dependent
critical masses in the presence of electromagnetism.
These must be determined nonperturbatively,
and various methods for doing so have been used in the literature.
One of these methods, proposed in Ref.~\cite{deDivitiis:2013xla},
can be implemented using partially quenched (PQ) \chpt, and thus
checked. This is done in App.~\ref{app:tune}.
We find that the method only provides one constraint on the up and
down critical masses and must be supplemented by an additional condition
in order to determine both.

Appendix~\ref{app:tune} requires results from a \chpt\ 
analysis of a theory with twisted nondegenerate charged quarks at nonzero
lattice spacing {\em and at nonvanishing $\thetaQCD$}. 
We provide such an analysis in a companion paper~\cite{chptforthetaqcd}.

\section{Power-counting and summary of previous work}
\label{sec:power}

In order to study the low-energy properties of LQCD, we must decide on
the relative importance of the competing effects.
The power counting that we adopt is
\begin{equation}
m \sim p^2 \sim a^2 \sim \alpEM > \epsilon_q^2 > m a\sim a^3\sim a \alpEM  ...
\,,
\label{eq:power}
\end{equation}
where $m$ represents either $m_u$ or $m_d$.
This is the power counting adopted in Ref.~\cite{Horkel:2014nha},
except that electromagnetic effects are now included.
This scheme only makes sense if discretization errors linear in
$a$ are absent, either because the action is improved or because
the $O(a)$ terms can be absorbed into a shift in the quark masses
(as is the case in W\chpt~\cite{Sharpe:1998xm}).

The explanation for the choice of leading order (LO) terms in this
power-counting is as follows.
Present simulations have $1/a\approx 3\;$GeV, 
and using this together with $\Lambda_{\rm QCD}\approx300\;$MeV
we find $a\Lambda_{\rm QCD}\approx 0.1$.
Thus second order discretization effects are of
relative size $(a\Lambda_{\rm QCD})^2\approx 0.01$.
This is comparable to $\alpEM$,
$m_u/\Lambda_{\rm QCD}$ and $m_d/\Lambda_{\rm QCD}$
(given that $m_u\approx 2.5\;$MeV
and $m_d\approx 5\;$MeV~\cite{Aoki:2013ldr,Agashe:2014kda}).
The results for the neutron-proton mass difference 
described in the Introduction are consistent with this power-counting
(using the fact that $m_u\!-\!m_d\sim m_u\sim m_d$).

The choice of $\epsilon_q^2$ as the dominant subleading contribution is less
obvious, and is discussed in some detail in Ref.~\cite{Horkel:2014nha}.
The essence of the argument is that, while the $\epsilon_q^2$ terms 
are not necessarily numerically larger than generic $m^2$ terms,
they give the leading 
contribution from quark mass differences to isospin breaking 
in the low-energy effective theory. For example, these contributions give
rise to the CP-violating phase in the continuum analysis.\footnote{%
A further justification for this choice, also discussed in 
Ref.~\cite{Horkel:2014nha}, is that in SU(3) \chpt\ such terms are
of LO, since they are proportional to $(m_u\!-\!m_d)^2/m_s$.}

In this note we keep only terms up to and including those
proportional to $\epsilon_q^2$, so that we have the leading order
term of each type.
We refer to this as working at  \LOP\,
indicating that it goes slightly beyond keeping only LO terms.

\bigskip

We now collect the relevant results from Ref.~\cite{Horkel:2014nha}
concerning the phase diagram of Wilson-like fermions in the presence
of nondegeneracy.
We work entirely in SU(2) W\chpt, in which the chiral field
is $\Sigma\in {\rm SU(2)}$.
The \LOP\ chiral Lagrangian for Wilson-like fermions
(whether improved or not) is
\begin{align}
\mathcal{L}_\chi &= \frac{f^2}{4}\tr\left[
\partial_\mu \Sigma \partial_\mu \Sigma^\dagger\right]
+ {\cal V}_\chi
\label{eq:Lchi}
\\
\mathcal{V}_\chi&= -\frac{f^2}{4} 
\tr({\chi}^\dagger \Sigma + \Sigma^\dagger {\chi})
- W' [\tr(\hat A^\dagger \Sigma +   \Sigma^\dagger \hat A)]^2
\nonumber\\
&\quad +\frac{\ell_7}{16}[\tr(\chi^\dagger \Sigma - \Sigma^\dagger \chi)]^2
\,,
\label{eq:Vchi}
\end{align}
where $\hat A=2 W_0 a \mathbb 1$ is the spurion field used to
introduce lattice artifacts.
This Lagrangian contains several LECs:
$f\approx 92\;$MeV and $B_0$ from LO continuum \chpt, 
$W_0$ and $W'$ introduced by disretization errors,
and $\ell_7$.
The latter, though of next-to-leading order (NLO) in
standard continuum power-counting, leads to
contributions proportional to $\epsilon_q^2$ and thus
we keep it in our LO+ calculation.
$\ell_7$ is not renormalized at one-loop order, and matching with
SU(3) \chpt\ leads to the estimate~\cite{Gasser:1984gg}
\begin{equation}
\ell_7=\frac{f^2}{8B_0m_s}\,,
\label{eq:ell7match}
\end{equation}
indicating that $\ell_7$ is positive.

The final ingredient in Eq.~(\ref{eq:Vchi}) is
$\chi=2 B_0 M$, which contains the mass matrix $M={\rm diag}(m_u,m_d)$,
with $m_{u,d}$ renormalized masses in a mass-independent scheme.
Since ${\cal L}_\chi$ is supposed to represent the long-distance physics
of a lattice simulation close to the chiral and continuum limits, to
use it we need to know the relationship between bare lattice masses
and the renormalized masses. This 
relationship is nontrivial when using nondegenerate quarks, and
is discussed in Appendix~\ref{app:renorm}.
This point was overlooked in Ref.~\cite{Horkel:2014nha}.

\begin{figure}[tb!]
\centering
\begin{subfigure}{0.49\textwidth}
\includegraphics[scale=.29]{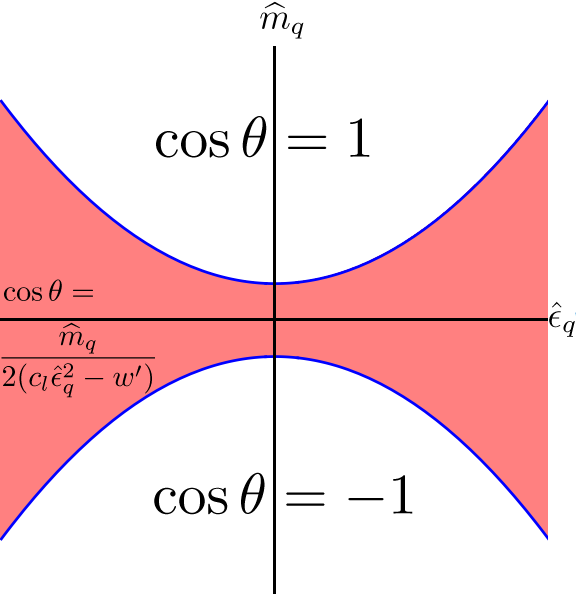}
\caption{\label{fig:Aokiold} Aoki scenario ($w'<0$).}
\end{subfigure}
\begin{subfigure}{0.49\textwidth}
\includegraphics[scale=.29]{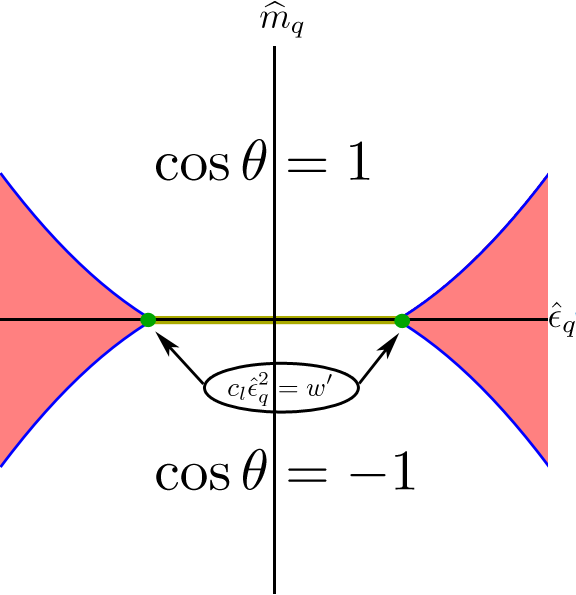}
\caption{\label{fig:Firstold} First-order scenario ($w'>0$).}
\end{subfigure}
\caption{
Phase diagrams from Ref.~\cite{Horkel:2014nha}
including effects of both discretization and nondegenerate quarks.
CP is violated in the (pink) shaded regions.
The (blue) lines at the boundaries of the shade regions
are second-order transitions (where the neutral pion mass vanishes), 
while the (yellow) line along the
$\epsilon_q$ axis joining the two shaded regions in (\ref{fig:Firstold})
is a line of first order transitions.
The analytic expression given for the shaded region in (\ref{fig:Aokiold})
holds also for that in (\ref{fig:Firstold}).
As discussed below in Sec.~\ref{subsec:Wilsonphase}, 
these phase diagrams apply also in the presence of electromagnetism.
}
\label{fig:phaseold}
\end{figure}

To determine the vacuum of the theory,
we must minimize the potential ${\cal V}_\chi$.
Writing $\langle\Sigma\rangle=e^{i\theta \hat n\cdot\vec \tau}$,
the potential becomes
\begin{equation}
{\mathcal{V}_\chi}
= - {f^2} \left( 
\mhat_q \cos{\theta} +c_\ell \ephat_q^2 n_3^2 \sin^2{\theta} 
+ w' \cos^2{\theta}\right)\,,
\label{eq:Vchi2}
\end{equation}
where 
\begin{multline}
\mhat_q= B_0 (m_u\!+\!m_d), \ \ephat_q=2 B_0\epsilon_q, 
\\ 
c_\ell=\frac{\ell_7}{f^2},  \ w'=\frac{64W'W_0^2 a^2}{f^2}.
\label{eq:mhatdef}
\end{multline}
Assuming $c_\ell > 0$ [based on the estimate (\ref{eq:ell7match})],
the resulting phase diagrams are shown in Fig.~\ref{fig:phaseold}.
The unshaded phases are continuum-like with $\lvert\cos\theta\rvert=1$.
The shaded (pink) phases violate CP with
\begin{equation}
|n_3|=1, \ \ \cos\theta=\frac{\mhat_q}{2(c_\ell \ephat_q^2-w')}.
\label{eq:condCP}
\end{equation}
The boundaries between continuum-like and CP-violating
phases lie along the lines $|\mhat_q| = 2 (c_\ell \ephat_q^2-w')$, 
and are second order transitions. 
The boundary between the two continuum-like phases with opposite $\cos\theta$
is a first order transition.
Within the continuum-like phases the pion masses are
\begin{equation}
m_{\pi^0}^2= |\mhat_q| - 2 (c_\ell \ephat_q^2-w')\,,
\ \
m_{\pi^\pm}^2= |\mhat_q| + 2 w'\,,
\label{eq:mpiWilson}
\end{equation}
while within the CP-violating phase
\begin{equation}
m_{\pi^0}^2=
2( c_\ell \ephat_q ^2  -w') \sin^2\theta\,,
\ \ 
m_{\pi^\pm}^2= 2c_\ell \ephat_q^2\,.
\label{eq:mpiWilsonCP}
\end{equation}
The neutral pion mass vanishes along the second order
transition lines.
Plots of these masses are given in Ref.~\cite{Horkel:2014nha}.

\section{Charged, nondegenerate Wilson quarks}
\label{sec:Wilson}

We now add electromagnetism, so that we are considering
Wilson fermions with charged, nondegenerate quarks.
Precisely how electromagnetism is added at the lattice level is not relevant;
all we need to know is that electromagnetic gauge invariance
is maintained by coupling to exact vector currents of the lattice theory.
We work here only at LO in $\alpEM$, which in terms of
Feynman diagrams means keeping only those with a single photon
propagator. 
We also work at infinite volume, thus avoiding the complications
of power-law volume dependence that occur in 
simulations~\cite{Hayakawa:2008an,Davoudi:2014qua,Borsanyi:2014jba}.

\subsection{Induced shifts in quark masses}
\label{subsec:Wilsonshift}

\begin{figure}
\begin{subfigure}{0.49\textwidth}
\includegraphics[width=\textwidth]{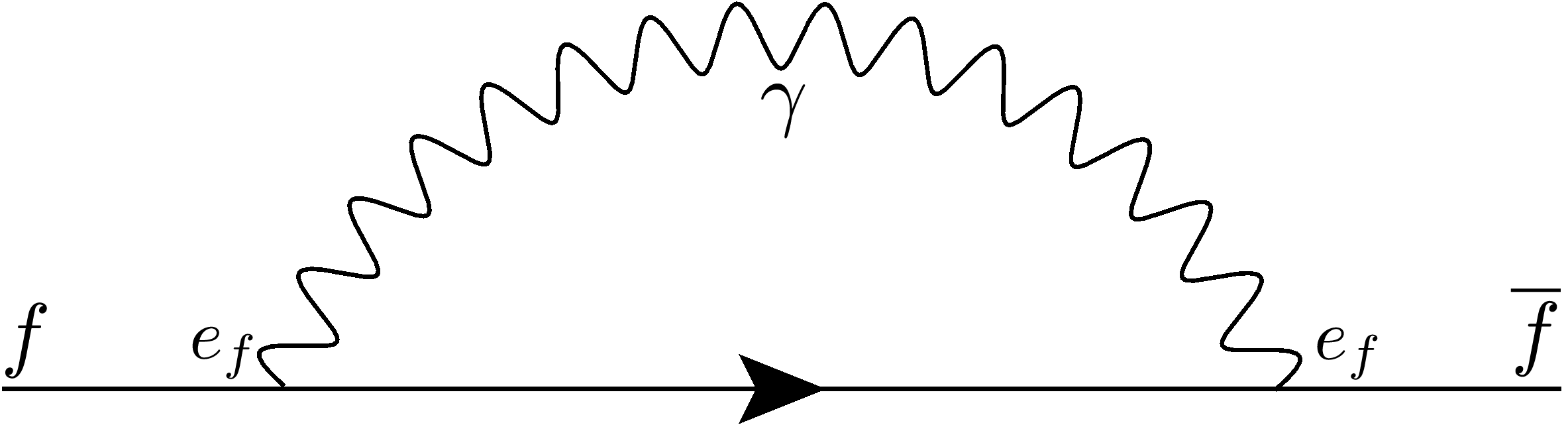}
\caption{ }
\end{subfigure}
\vspace{0.2in}

\begin{subfigure}{0.49\textwidth}
\includegraphics[width=\textwidth]{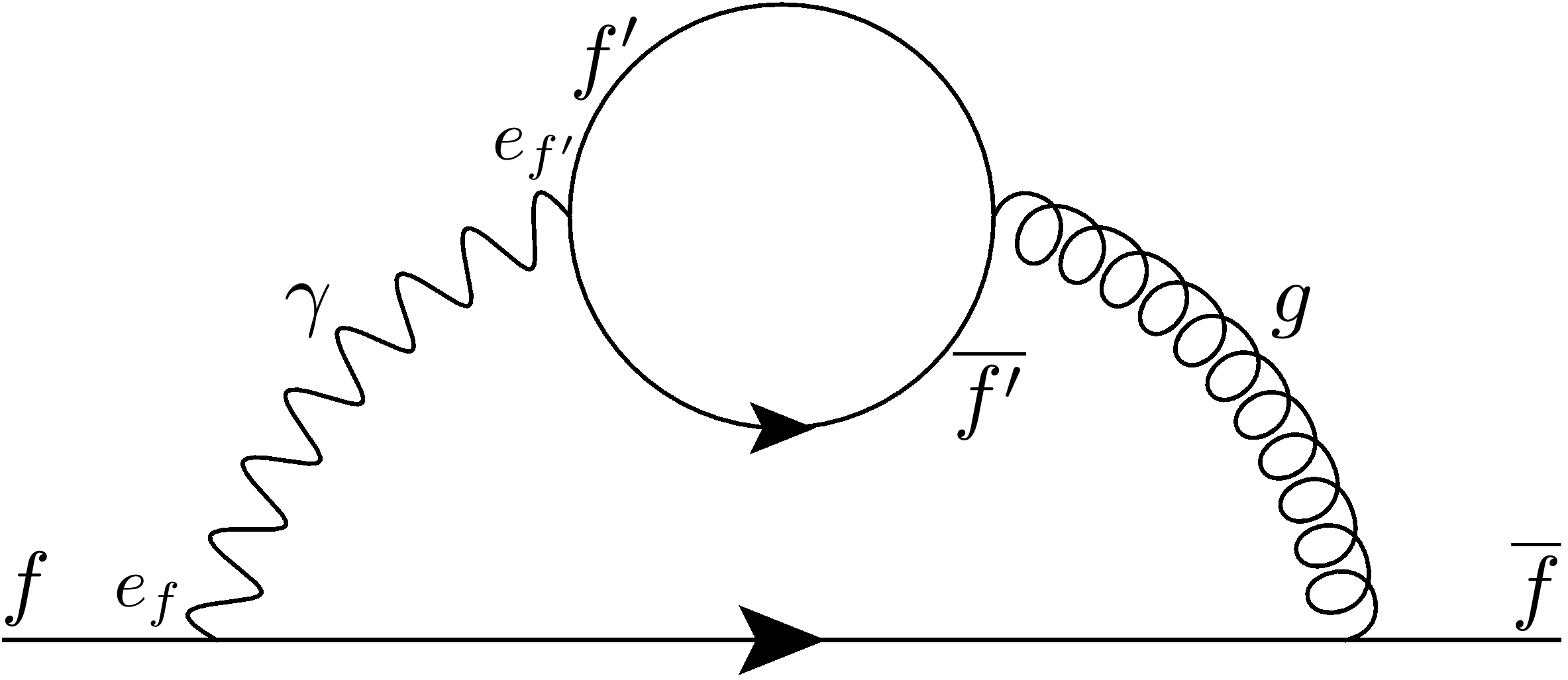}
\caption{ }
\end{subfigure}
\vspace{0.2in}

\begin{subfigure}{0.49\textwidth}
\includegraphics[width=\textwidth]{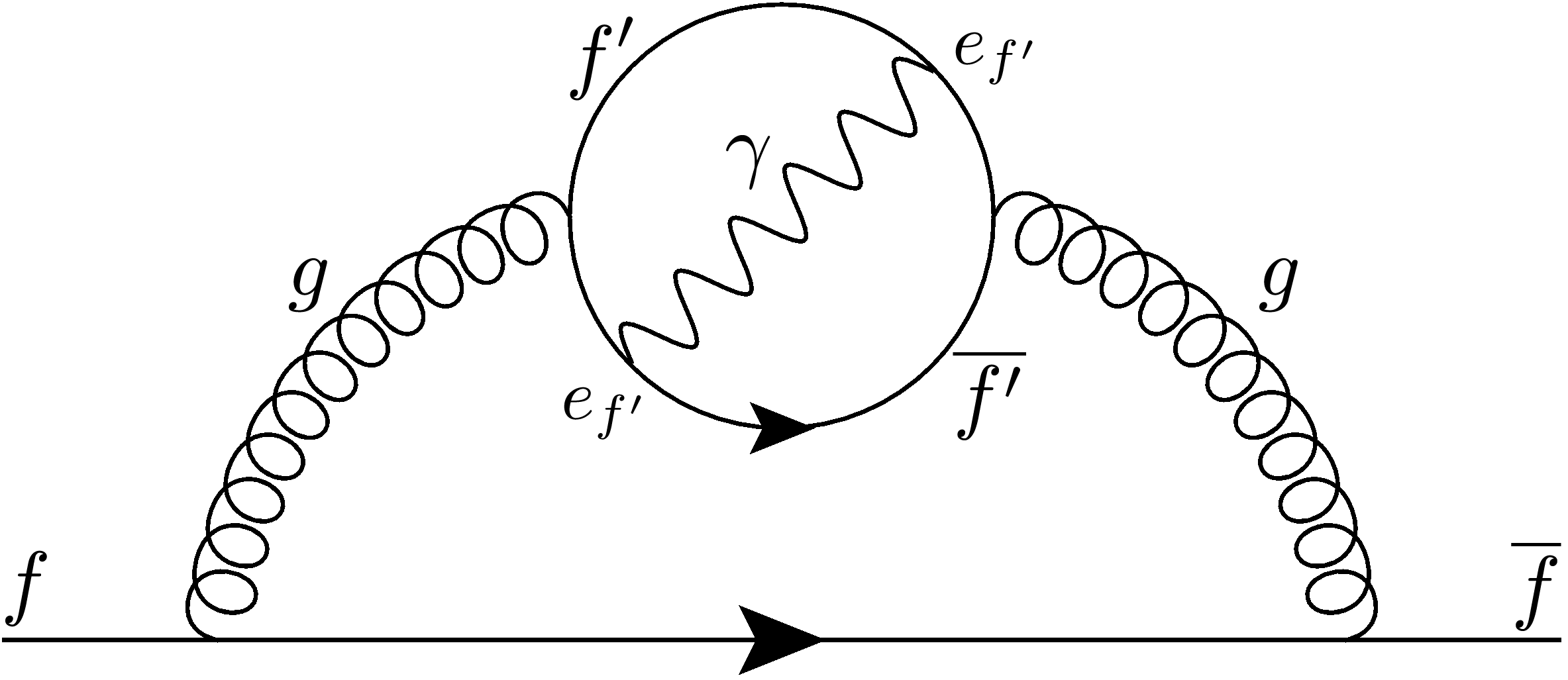}
\caption{ }
\end{subfigure}
\vspace{0.2in}

\caption{\label{fig:feyn} Examples of
LO contributions from electromagnetism to quark self-energies.
Diagrams with additional gluons and quark loops are not shown.
These three types of diagram lead, respectively, 
to the three operators listed in Eq.~\ref{eq:EMops}.
Only the first operator is present in the ``electroquenched'' 
approximation.
}
\end{figure}

The dominant effect of electromagnetism is a charge dependent
shift in the critical mass, as noted in
Refs.~\cite{Portelli:2010yn,deDivitiis:2013xla,Borsanyi:2014jba}.
Here we discuss this shift from the viewpoint of the Symanzik low-energy 
effective Lagrangian~\cite{Symanzik:1983dc,Symanzik:1983gh}.
It arises from QCD self-energy diagrams in which one 
of the gluons is replaced by a photon,
and leads to the appearance of the operators
\begin{align}
(a) &\frac{\alpEM}{a} (\sum_f e_f^2 \overline f f)\,,
\nonumber \\
(b) &\frac{\alpEM}{a} (\sum_{f'} e_{f'}) \sum_{f} e_{f} \overline f f\,,
\nonumber\\
(c) &\frac{\alpEM}{a} \sum_{f'} (e_{f'}^2) \sum_{f} \overline f f\,,
\label{eq:EMops}
\end{align}
where $f=u,d$, $e_u=2/3$ and $e_d=-1/3$. 
Examples of the corresponding Feynman diagrams are shown in Fig.~\ref{fig:feyn}

These operators are allowed because electromagnetism breaks isospin, while
Wilson fermions violate chiral symmetries.
Their contributions are smaller than those of the $\sum_f\bar f f/a$ operator that leads to the
dominant shift in the critical mass.
However, because $\alpEM\sim a^2 \sim m$ in our power-counting, 
$\alpEM/a$ effects are proportional to $a \sim m^{1/2}$, 
and thus dominate over physical quark masses.
They must therefore be removed by appropriate tuning of the bare masses.
Since the combined effect of the three operators is independent
${\cal O}(\alpEM/a)$ shifts in $m_u$ and $m_d$,
removing these shifts requires independent tuning of the
$u$ and $d$ critical masses.

Different methods for doing this tuning have been used in the literature.
The most straightforward, used in Ref.~\cite{Borsanyi:2014jba},
is to determine the bare quark masses directly by enforcing that
an appropriate subset of hadron masses agree with their
experimental values (keeping all isospin breaking effects).
This avoids the need to directly determine the critical masses,
but is the most challenging numerically.
An alternative approach, proposed in
Ref.~\cite{deDivitiis:2013xla}, makes use of a partially-quenched
extension of the theory.
In Appendix~\ref{app:tune} we check this method by showing
how it can be implemented in \chpt.
We find that it cannot determine both critical masses,
but instead only provides a single constraint between them.
We then introduce an additional tuning criterion which,
together with that of Ref.~\cite{deDivitiis:2013xla},
does allow both critical masses to be determined.

For the rest of the main text, we assume that
the charge-dependent critical masses have been determined
in some manner, such that ${\cal O}(\alpEM/a)$ 
self-energy effects can be ignored.
This leaves electromagnetic corrections proportional to $\alpEM$,
which we must keep in our power counting, as well as higher-order
effects proportional to $\alpEM\times m$ etc., which we can ignore.

Examples of the latter effects are the bilinears 
\begin{equation}
\alpEM \sum_f e_f^2 m_f \bar f f
\ \ {\rm and}\ \ 
\alpEM \sum_f e_f^2 \bar f \slashed{D} f\,.
\end{equation}
These arise as ${\cal O}(am)$ corrections to the operators of Eq.~(\ref{eq:EMops}),
and are also present directly in the continuum theory.
We stress that, in the Symanzik Lagrangian, one has no dimensionful parameters
other than $m$ and $1/a$, so bilinears proportional to $\alpEM \Lambda_{\rm QCD}$
are not allowed. Factors of $\Lambda_{\rm QCD}$ arise when we move from
the Symanzik Lagrangian to \chpt.

The only effect of electromagnetism that is
simply proportional to $\alpEM$---and thus
of LO in our power counting---is that
arising from one photon exchange between electromagnetic currents. 
This is a continuum effect, long studied in \chpt.
It leads to he following additional term in the
chiral potential~\cite{Gupta:1984tb,Ecker:1988te}:\footnote{%
Contributions from the isoscalar part of the photon coupling lead
to the same form but with one or both $\tau_3$'s replaced by
identity matrices. In either case the contribution reduces to an
uninteresting constant, and is thus not included in ${\cal V}_{\rm EM}$.}
\begin{equation}
{\cal V}_\EM = - \frac{f^2}4 \cEM \tr(\Sigma \tau_3 \Sigma^\dagger \tau_3)\,.
\label{eq:VEM}
\end{equation}
Here $\cEM$ is an unknown coefficient proportional to $\alpEM$.
All that is known about $\cEM$ is that it is positive~\cite{Witten:1983ut}.

\subsection{Phase diagram and pion masses}
\label{subsec:Wilsonphase}

The competition between electromagnetic effects and discretization errors
for two {\em degenerate} Wilson fermions has been previously analyzed in 
Ref.~\cite{Golterman:2014yha}.
Here we add in the effects of nondegeneracy.
This turns out to be very simple. Using the SU(2) identity
\begin{equation}
4 \tr(\Sigma \tau_3 \Sigma^\dagger \tau_3)
= \left[\tr(\Sigma\!+\!\Sigma^\dagger)\right]^2
- \left[\tr([\Sigma\!-\!\Sigma^\dagger]\tau_3)\right]^2
-8,
\end{equation}
together with
\begin{equation}
\chi = \mhat_q {\mathbb 1} + \ephat_q \tau_3\,,
\end{equation}
we find that ${\cal V}_\EM$ can be absorbed into ${\cal V}_\chi$
[given in Eqs.~(\ref{eq:Vchi}) and (\ref{eq:Vchi2})]
by changing the existing constants as
\begin{equation}
w' \longrightarrow w' + \cEM\,,
\ \ {\rm and}\ \ 
c_\ell \ephat_q^2 \longrightarrow c_\ell \ephat_q^2+\cEM\,.
\label{eq:constantshift}
\end{equation}
This allows us to determine the phase diagram and pion masses
directly from the results presented in the previous section.\footnote{%
For $\ephat_q=0$ our results are in complete agreement
with those of Ref.~\cite{Golterman:2014yha}.}

We first observe that, at the order we work, the phase diagram is {\em
unchanged by the inclusion of EM}---the results in 
Fig.~\ref{fig:phaseold} still hold.
This can be seen from the form of the potential in
Eq.~(\ref{eq:Vchi2}), which, since $|n_3|=1$, depends only
on $c_\ell \ephat_q^2-w'$. This combination is, however,
unaffected by the shifts of Eq.~(\ref{eq:constantshift})
and so the phase boundaries and values of $\theta$
throughout the phase plane are also unchanged.

Similarly, from Eqs.~(\ref{eq:mpiWilson}) and
(\ref{eq:mpiWilsonCP}) we see that the neutral pion masses
are unchanged throughout the phase plane.
In particular, the second-order phase boundaries are
(as expected) lines along which the neutral pion is massless.

\begin{figure}
\centering
\begin{subfigure}{0.49\textwidth}
\includegraphics[scale=.5]{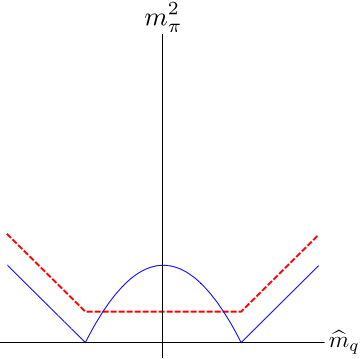}
\caption{Aoki scenario with $w' < -\cEM$}
\end{subfigure}

 \begin{subfigure}{0.49\textwidth}
\includegraphics[scale=.5]{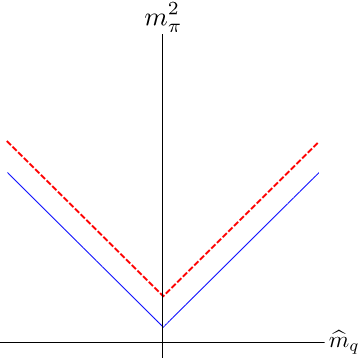}
\caption{First-order scenario with $w'>c_\ell\ephat_q^2$}
\end{subfigure}

 \begin{subfigure}{0.49\textwidth}
\includegraphics[scale=.5]{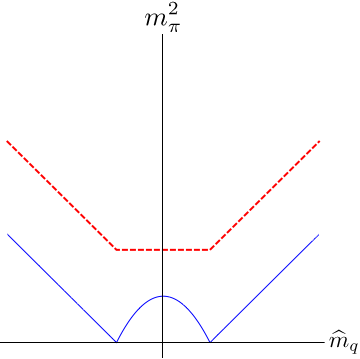}
\caption{First-order or Aoki scenario with $-\cEM < w' < c_\ell\ephat_q^2$}
\end{subfigure}

\caption{\label{fig:untwistedpions} 
Pion masses for nondegenerate untwisted Wilson fermions including electromagnetism.
The three possible behaviors along
vertical slices through phase diagrams of Fig.~\ref{fig:phaseold} are shown.
Solid (blue) lines show $m_{\pi^0}^2$, while dashed (red) lines show $m_{\pi^\pm}^2$.
Expressions for masses are given in the text.}
\end{figure}

The only change caused by electromagnetism is to the charged pion masses, 
which are increased by the same amount throughout the phase plane:
\begin{equation}
m_{\pi^\pm}^2 \longrightarrow m_{\pi^\pm}^2 + 2 \cEM\,.
\label{eq:globalshiftmpiplus}
\end{equation}
One  implication is that, for $\ephat_q=0$, the charged pions
are no longer massless within the Aoki phase (if present).
This is because they are no longer Goldstone bosons, as the
flavor symmetry is explicitly broken by electromagnetism.
Also, as noted in Ref.~\cite{Golterman:2014yha}, the charged
pion can be lighter than the neutral one inside the CP-violating
phases. This is not inconsistent with Witten's 
identity~\cite{Witten:1983ut} because the latter did not account
for discretization effects.
Plots of the pion masses are shown in Fig.~\ref{fig:untwistedpions}.

It is perhaps surprising that electromagnetism, which contributes at LO
in our power-counting, has no effect on the phase diagram,
whereas the subleading contributions proportional to $\ephat_q^2$
have a significant impact.
We can understand this by noting that the CP-violating
phase is characterized by a neutral pion condensate, 
which remains uncoupled to the photon 
until higher order in \chpt\ (where form factors enter).

The implications of these results for practical simulations
(such as those of Ref.~\cite{Borsanyi:2014jba})
are unchanged from the discussion in Ref.~\cite{Horkel:2014nha}.
In particular, for the Aoki scenario ($w'<0$) discretization
effects move the CP-violating phase closer to the physical point
than for degenerate quarks, so one must beware of simulating
too close to this transition.

\section{Nondegeneracy, electromagnetism and twist}
\label{sec:twistEM}

When using twisted-mass fermions one must decide 
on the relative orientation in isospin space both
of the twist and the isospin-breaking induced
by quark mass differences and electromagnetism.
In the absence of electromagnetism, the standard
choice is to align these two effects in orthogonal directions.
For example, one usually takes
$\tau_3$ for isospin-breaking, as in the continuum,
while twisting in the $\tau_1$ direction.\footnote{%
Any linear combination of $\tau_1$ and $\tau_2$ is equivalent;
$\tau_1$ is the standard choice.}
This is the choice used in simulations of the strange-charm
sector using twisted-mass fermions~\cite{Baron:2010th}.
It ensures that the fermion determinant is real, and
(subject to some conditions) positive~\cite{Frezzotti:2003xj}.
This was the choice whose phase structure
we determined using W\chpt\ in Ref.~\cite{Horkel:2014nha}.

This approach does not, however, allow for the inclusion of
electromagnetism. One problem is apparent already in the
continuum limit, where the twisted-mass quark 
action is (in the ``twisted'' basis)~\cite{Frezzotti:2000nk}
\begin{equation}
\overline\psi (\slashed{D} + m_q c_\omega
+ i\gamma_5\tau_1 m_q s_\omega +\epsilon_q \tau_3)\psi
\,.
\label{eq:continuumtau1twist}
\end{equation}
Here $\slashed{D}$ is the gluonic covariant derivative,
$m_q$ is the average quark mass, and
$\omega$ the twist angle with $c_\omega=\cos\omega$ and
$s_\omega=\sin\omega$.
This action is not invariant under flavor rotations in the
$\tau_3$ direction, so there is no conserved vector current
to which the photon can couple. In other words, there is no global
flavor transformation available to gauge.

To avoid this problem, we recall that twisting is, in the
continuum, simply a nonanomalous change of variables that
does not effect physical quantities. Thus we should start
with the standard action including electromagnetism
\begin{equation}
\overline\psi(\slashed{D} 
- i e \slashed{A} Q +
m_q + \epsilon_q \tau_3) \psi\,,
\end{equation}
with $A_\mu$ the photon field coupling via the charge matrix 
\begin{equation}
Q = \frac16 {\mathbb 1} +  \frac12 \tau_3
\,,
\end{equation}
and then perform a chiral twist
\begin{equation}
\psi \longrightarrow e^{i\omega \gamma_5 \tau_1/2}\psi,
\ \
\overline\psi \longrightarrow \overline\psi e^{i\omega \gamma_5 \tau_1/2}
\,.
\end{equation}
This leads to the quark action of Eq.~(\ref{eq:continuumtau1twist})
with the addition of the photon coupling
\begin{equation}
\overline\psi \slashed{A}
\left[\frac16 {\mathbb 1}
+ \frac12 (c_\omega \tau_3 - s_\omega \tau_2 \gamma_5)\right]
\psi
\,.
\label{eq:photontau1twist}
\end{equation}
Thus the photon couples to a linear combination of 
vector currents and to an
axial current in the $\tau_2$ direction.
In the continuum, this combination is conserved
[given the twisted mass matrix of Eq.~(\ref{eq:continuumtau1twist})]
and the action remains gauge invariant.

We conclude that the correct fermion action to discretize
is the sum of 
Eqs.~(\ref{eq:continuumtau1twist}) and (\ref{eq:photontau1twist}).
This, however, is not possible in a gauge invariant way
using Wilson's lattice derivative (except for $s_\omega=0$).
The Wilson term breaks all axial symmetries, so the $\tau_2\gamma_5$
part of the photon coupling is to a lattice current that is
not conserved.

To avoid this problem, and obtain a discretized twisted theory that
maintains gauge invariance, one needs to twist in a direction
that leaves the photon coupling to a conserved current.
The only choice is to twist in the $\tau_3$ direction.
Then the twisted form of the continuum Lagrangian is
\begin{equation}
\overline\psi (\slashed{D} -i e \slashed{A} Q
+ m_q c_\omega + \tau_3\epsilon_q c_\omega
+ i\gamma_5\tau_3 m_q s_\omega +i\gamma_5 \epsilon_q s_\omega)\psi
\,.
\label{eq:tau3twist}
\end{equation}
This is discretized by adding the standard Wilson term.
Since the photon is coupled to vector currents that
are exact symmetries of both the Wilson term and the full
mass matrix, gauge invariance is retained.

This form of the twisted isospin-violating action (with $\omega=\pi/2$) is 
used in the recent work of Refs.~\cite{deDivitiis:2011eh,deDivitiis:2013xla}.
It has one major practical disadvantage---the quark
determinant is complex for nonzero twist.
This is true for nondegenerate masses alone, as explained
in Ref.~\cite{Walkerloud:2005bt}. Adding electromagnetism only makes the
problem worse, since at the least it induces further
nondegeneracy in the masses.
Because the action is complex, direct simulation with present
fermion algorithms is challenging.
This problem is avoided in Refs.~\cite{deDivitiis:2011eh,deDivitiis:2013xla}
by doing a perturbative expansion in powers of $\epsilon_q$ and
$\alpEM$. The expectation values are then evaluated in the theory
with no isospin breaking, for which the fermion determinant with twisting
is real and positive.

In the following section we study the phase diagram of the theory
with the discretized form of the Lagrangian (\ref{eq:tau3twist}).
To our knowledge, this form of the twisted theory has not
previously been studied in W\chpt\ either with nondegeneracy
alone or with electromagnetism.

\section{\chpt\ for charged, nondegenerate quarks with a $\tau_3$ twist}
\label{sec:twistchiral}

The conclusion of the previous section is that the
twisted-mass theory whose phase diagram is of interest is that
with lattice fermion Lagrangian
\begin{equation}
\overline\psi_L\left[
{D}_W + m_{0} + \tau_3 \epsilon_{0}
+ i \gamma_5 \tau_3 \mu_0  + i \gamma_5 \eta_{0}\right]
\psi_L\,.
\label{eq:twistedlatticeaction}
\end{equation}
$\psi_L$ is a lattice fermion field
and  $D_W$ the lattice Dirac operator 
including the Wilson term (and possibly improved).
$D_W$ is coupled to both gluons and photons, with the
latter coupling to the $\tau_3$ vector current.
The action differs from that considered (implicitly)
in Sec.~\ref{sec:Wilson} only by the addition of
the two mass parameters $\mu_0$ and $\eta_{0}$.

The four bare mass parameters in (\ref{eq:twistedlatticeaction})
are related in the continuum limit to 
the renormalized up and down masses, the twist angle
(which is a redundant parameter) and the QCD theta angle,
$\theta_\QCD$. 
The aim is to tune the bare parameters so that the dimension 4
part of the quark contribution to the Symanzik
effective Lagrangian is given by
Eq.~(\ref{eq:tau3twist}) with the desired physical
quark masses, for some choice of $\omega$.
As for untwisted Wilson fermions 
the dominant effect of electromagnetism is to cause
separate ${\cal O}(\alpEM/a)$ shifts in the (untwisted) 
up and down masses.
These shifts depend on twisted masses only at quadratic order, so that,
to the order we work, they are identical to those for Wilson fermions.
They can be determined by the methods discussed in 
Sec.~\ref{subsec:Wilsonshift} and Appendix~\ref{app:tune}.
They are equivalent to independent shifts in $m_0$ and $\epsilon_{0}$.

After the additive shift in $m_0$ and $\epsilon_{0}$,
all four masses in (\ref{eq:twistedlatticeaction})
must be multiplicatively renormalized in order to be
related to the continuum masses in Eq.~(\ref{eq:tau3twist}).
As discussed in Appendix~\ref{app:renorm},
this requires different renormalization factors for all four masses.
We assume here that these renormalizations have been carried out,
so that the dimension four term in the Symanzik effective Lagrangian is
given by Eq.~(\ref{eq:tau3twist}) and described by the three parameters
$m_q$, $\epsilon_q$ and $\omega$.

We stress that this tuning and renormalization must be carried out
with sufficient accuracy. If not, instead of Eq.~(\ref{eq:tau3twist}), 
one ends up with a similar form having {\em different} twist angles 
for the $m_q$ and $\epsilon_q$ parts.
The parity-odd parts can then only be removed by a combined
flavor nonsinglet and flavor singlet twist. Since the latter is anomalous,
this corresponds to a theory with nondegenerate quark masses, 
electromagnetism,
a twist angle {\em and} a nonvanishing $\theta_\QCD$.
In other words, the theory not only has the unphysical parity 
violation due to twisting (which can be rotated away in the continuum limit) 
but also the physical parity violation induced by $\theta_\QCD$.
Indeed, to analyze the tuning in \chpt\ one needs to include
a nonvanishing $\theta_\QCD$, an analysis we carry out in
a companion paper~\cite{chptforthetaqcd}.

Assuming that the dimension-four quark Lagrangian is
Eq.~(\ref{eq:tau3twist}), 
we next investigate which higher-dimension operators
are introduced into the Symanzik Lagrangian by twisting.
Those operators present for Wilson fermions remain, but, 
as discussed in Sec.~\ref{sec:Wilson},
are all of higher order than we consider.
The dominant operators introduced by twisting will violate parity,
because they are linear in the parity-violating mass terms $\mu_0$ and $\eta_0$.
Examples of the new operators are\footnote{%
The first of these corresponds to an induced value of 
$\theta_\QCD$ proportional to $a\eta_{0}$. This is one way
of seeing that the lattice action (\ref{eq:twistedlatticeaction})
leads to a complex fermion determinant.}
\begin{equation}
a \eta_{0} G_{\mu\nu}\widetilde G_{\mu\nu}\,,
\ 
a \eta_{0} \overline\psi \widetilde G_{\mu\nu}\sigma_{\mu\nu} \psi\,,
\ {\rm and}\ 
a \mu_{0} \overline\psi \tau_3 \widetilde G_{\mu\nu}\sigma_{\mu\nu} \psi\,.
\label{eq:PVops}
\end{equation}
Since we generically treat $a m$ terms as being beyond \LOP\
[see Eq.~(\ref{eq:power})], we should be able to ignore these operators.
However, because $\eta_0\sim \epsilon_q$ and we are treating $\epsilon_q$ as somewhat enhanced,
one might be concerned about dropping $a \eta_0$ terms.
In fact, the $a \eta_0$ operators in (\ref{eq:PVops}), when matched into \chpt, pick up an additional
factor of $m$ or $p^2$, and thus are unambiguously suppressed.
The reason for the extra factors is 
that the LO representation of a flavor-singlet pseudoscalar
in \chpt, $\tr(\Sigma-\Sigma^\dagger)$, vanishes identically.
For the induced $\theta_\QCD$ term, one can also see this result by
noting that it can be rotated into the isosinglet mass term, leading to
a contribution proportional to $m \theta_\QCD\sim a \epsilon m$.

Proceeding in this fashion, we find that all other new
operators induced by the parity-breaking masses are beyond
\LOP\ in our power counting.
Thus, once the requisite tuning has been done, the \LOP\ chiral effective
theory for $\tau_3$ twisted fermions with isospin breaking is
given by the same result as for Wilson fermions, i.e.
\begin{align}
\frac{f^2}{4}\tr\left[
\partial_\mu \Sigma \partial_\mu \Sigma^\dagger\right]
+ {\cal V}_\chi + {\cal V}_\EM
\end{align}
[see Eqs.~(\ref{eq:Vchi}) and (\ref{eq:VEM})],
except that the quark mass matrix is now twisted
\begin{align}
\chi 
&= (\mhat_q +\ephat_q\tau_3) e^{i\omega\tau_3} \,.
\label{eq:chitwisted}
\end{align}
We analyze the phase structure of this chiral theory in the
next two subsections.

\subsection{Phase diagram and pion masses at maximal $\tau_3$ twist}
\label{subsec:maxtwist}

We begin working at maximal $\tau_3$ twist, which is the choice
used in Refs.~\cite{deDivitiis:2011eh,deDivitiis:2013xla}.
In this case
\begin{equation}
\chi 
= i\mhat_q\tau_3 + i\ephat_q \,,
\label{eq:chimaxtwist}
\end{equation}
and the chiral potential becomes
\begin{align}
-\frac{\mathcal{V}_{\chi+\EM}}{f^2}
&= 
\mhat_q n_3 \sin{\theta} - (c_\ell \ephat_q^2 + w')\sin^2\theta 
\nonumber\\
&\qquad\qquad +\cEM (\cos^2\theta+n_3^2\sin^2\theta)\,,
\label{eq:Vchi3}
\end{align}
up to an irrelevant constant.
Since $\cEM>0$, the right-hand side is maximized always with $|n_3|=1$, and
we see that the $\cEM$ term becomes a constant.
Thus, once again, electromagnetism has no impact on the phase diagram.
We also see that the effect of nondegeneracy can be deduced from
the results for the degenerate case
(studied in Refs.~\cite{Munster:2004am,Scorzato:2004da,Sharpe:2004ps})
simply by shifting $w'$.

\begin{figure}[tb!]
\centering
\begin{subfigure}{0.49\textwidth}
\includegraphics[width=0.75\textwidth]{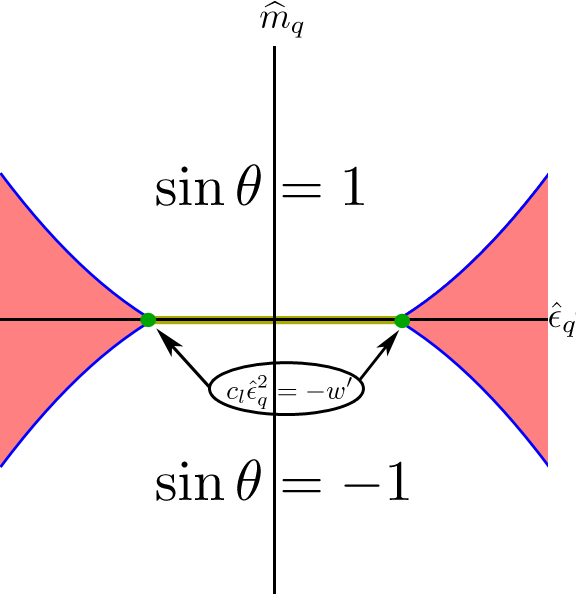}
\caption{\label{fig:Aokitwist} Aoki scenario ($w'<0$).}
\end{subfigure}
~
\begin{subfigure}{0.49\textwidth}
\includegraphics[width=0.75\textwidth]{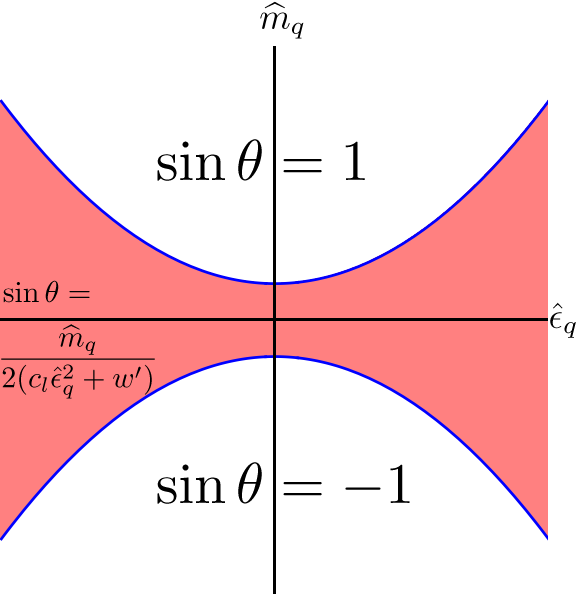}
\caption{\label{fig:Firsttwist} First-order scenario ($w'>0$).}
\end{subfigure}
\caption{\centering Phase diagrams including effects of 
discretization and nondegeneracy for maximally $\tau_3$-twisted quarks.
Electromagnetism has no impact on the phase diagram.
Notation as in Fig.~\ref{fig:phaseold}. The neutral pion is
massless along the second-order phase boundary
between shaded (CP-violating) and unshaded phases.
}
\label{fig:phasetwist}
\end{figure}

The resulting phase diagrams are shown in Fig.~\ref{fig:phasetwist}.
Comparing to the untwisted results of Fig.~\ref{fig:phaseold},
we see that the role of the Aoki and first-order scenarios has
interchanged. Without loss of generality, we can take $n_3=1$
throughout the phase plane. 
Then, in the continuum-like (unshaded) phases we have 
$\sin\theta={\rm sign}(\mhat_q)$, corresponding to the condensate
aligning or antialigning with the applied twist.
Second order transitions occur at $|\mhat_q| = 2(w' + c_\ell \ephat_q^2)$.
For smaller values of $|\mhat_q|$ the condensate angle
is $\sin\theta=\mhat_q/(2[w'+c_\ell\ephat_q^2])$,
with two degenerate minima having opposite signs of $\cos\theta$.
If one switches to the ``physical basis'' in which the twist is 
put on the Wilson term, then one finds that this phase violates
CP, just as in the Wilson case.

These results differ significantly from the phase structure
for nondegenerate quarks with a maximal $\tau_1$ twist,
shown in Fig. 3 of Ref.~\cite{Horkel:2014nha}.
In particular, an additional phase found for $w'>0$ with a $\tau_1$ twist
is absent here. We stress again that 
only the theory with a $\tau_3$ twist, i.e. that discussed here,
can incorporate electromagnetism.

\begin{figure}
\centering
\begin{subfigure}{0.49\textwidth}
\includegraphics[scale=.5]{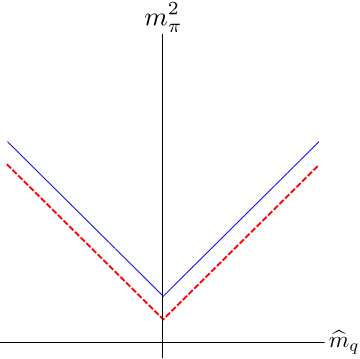}
\caption{Aoki scenario with $c_\ell\ephat_q^2+w' < -\cEM< 0$}
\end{subfigure}

 \begin{subfigure}{0.49\textwidth}
\includegraphics[scale=.5]{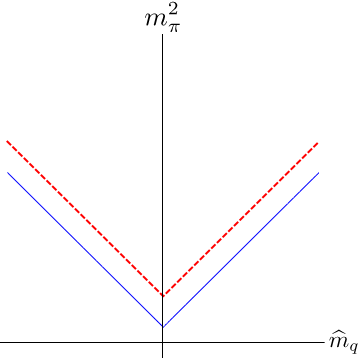}
\caption{Aoki scenario with $-\cEM < c_\ell\ephat_q^2+w' < 0$}
\end{subfigure}

 \begin{subfigure}{0.49\textwidth}
\includegraphics[scale=.5]{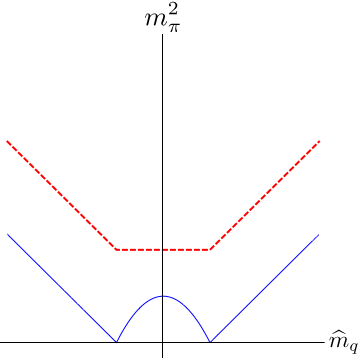}
\caption{Aoki or first-order scenario with $c_\ell\ephat_q^2+w'>0$}
\end{subfigure}

\caption{\label{fig:twistedpions} 
Pion masses for nondegenerate maximally $\tau_3$-twisted 
fermions including electromagnetism.
The three possible behaviors along
vertical slices through phase diagrams of Fig.~\ref{fig:phasetwist} are shown.
Solid (blue) lines show $m_{\pi^0}^2$, while dashed (red) lines show $m_{\pi^\pm}^2$.
Expressions for masses are given in the text.
}
\end{figure}

For the pion masses we find the following results.
Within the continuum-like phases we have
\begin{equation}
m_{\pi^0}^2= |\mhat_q| - 2 (c_\ell \ephat_q^2+w')\,,
\ \
m_{\pi^\pm}^2= |\mhat_q| + 2 \cEM\,,
\label{eq:mpitwist}
\end{equation}
while within the CP-violating phase
\begin{equation}
m_{\pi^0}^2=
2( c_\ell \ephat_q ^2  +w') \cos^2\theta\,,
\ \ 
m_{\pi^\pm}^2= 2(c_\ell \ephat_q^2+w'+\cEM) \,.
\label{eq:mpitwistCP}
\end{equation}
As expected, only the charged pion masses are affected by
electromagnetism. 
Plots of these results along vertical slices through the phase
diagram are shown in Fig.~\ref{fig:twistedpions}.

It is interesting to compare to the results with untwisted fermions,
which are given in Eqs.~(\ref{eq:mpiWilson}) and (\ref{eq:mpiWilsonCP})
together with the shift (\ref{eq:globalshiftmpiplus}) of $m_{\pi^\pm}^2$ by $2\cEM$ induced by
electromagnetism.
We see that the neutral pion mass differs only 
by the change of sign of $w'$
(which also implies the interchange $\sin\theta \leftrightarrow\cos\theta$).
This means that the results in the two scenarios interchange exactly.
For the charged pion masses, apart from the interchange of scenarios
there are also overall shifts proportional to $w'$.

The implications of these results for present simulations are as follows.
If one could simulate the theory directly (somehow dealing with the fact that the action is
complex) then one would need to avoid working in or near the CP-violating phase. This is now
more difficult in the first-order scenario than the Aoki scenario---opposite to the
situation with untwisted Wilson fermions. This qualitative result is the same as
for $\tau_1$ twisting (without electromagnetism), 
although the area taken up by unphysical phases 
is larger in that case~\cite{Horkel:2014nha}.
As noted above, actual simulations done to date at maximal twist use perturbation
theory in $\ephat_q$ and $\alpEM$, and so evaluate all
expectation values in the theory with $\ephat_q=\alpEM=0$.
Clearly, if $w'>0$, these simulations must be careful to have $\hat m_q$ large enough
to avoid the CP-violating phase.\footnote{%
In addition, if these simulations are done close 
to the onset of the CP-violating phase,
one would expect the expansion in $\ephat_q$ to be poorly convergent.
This is probably not a problem for the method of Ref.~\cite{Frezzotti:2000nk}, 
however,
since they take the continuum limit of the term linear in $\ephat_q$, and in this
limit $w'=0$ and the lattice artifacts discussed here vanish.}

\subsection{Nonmaximal $\tau_3$ twist}
\label{subsec:nonmaxtwist}

We have also investigated the phase structure for
general $\tau_3$ twist, i.e. nonvanishing and nonmaximal.
One motivation for doing so is that twisted-mass simulations 
cannot achieve exactly maximal twist; another 
is to see how the phase diagrams of Fig.~\ref{fig:phaseold}
change into those of Fig.~\ref{fig:phasetwist}.

Expressions are simplified if we define $\theta$ relative
to a twist $\omega$, i.e. if we use
\begin{equation}
\langle\Sigma\rangle = e^{i\omega\tau_3/2}
e^{i\theta \hat n\cdot \vec \tau}
e^{i\omega\tau_3/2}\,.
\label{eq:sigmaparam}
\end{equation}
Then we find (dropping constants)\footnote{%
At $\omega=\pi/2$ this should agree with Eq.~(\ref{eq:Vchi3}),
and it does once the different definitions of $\theta$ and $\hat n$
are taken into account.}
\begin{align}
-\frac{\cal V}{f^2}
&=
\mhat_q \cos\theta + c_\ell \ephat_q^2 n_3^2\sin^2\theta
\nonumber\\
&\quad + w' (\cos\theta\cos\omega-n_3 \sin\theta\sin\omega)^2
\nonumber\\
&\quad + \cEM (\cos^2\theta+ n_3^2 \sin^2\theta)
\,.
\label{eq:Varb}
\end{align}
This is not amenable to simple analytic extremization, and we
have used a mix of analytic and numerical methods.
One can show analytically that the minima always occur at $|n_3|=1$.
This implies that, once again, the electromagnetism does not
play a role in determining the phase structure.

The sign of $n_3$ can always be absorbed into $\theta$,
so we can again choose $n_3=1$ without loss of generality.
The potential can then be written (up to $\theta$-independent terms) as
\begin{align}
-\frac{\cal V}{f^2}\Bigg|_{n_3=1}
&=
\mhat_q \cos\theta
+ \cos^2\theta \left[-c_\ell \ephat_q^2 + w' \cos(2\omega)\right]
\nonumber\\
&\quad - \frac{w'}2 \sin(2\theta)\sin(2\omega)
\,.
\end{align}
A numerical investigation of this potential finds that,
for nonextremal $\omega$, and for all nonzero $w'$,
there is a first-order transition as $\mhat_q$ passes through zero,
irrespective of the value of $\ephat_q$.
At this transition $\theta$ jumps from $\pi/2 - \delta$ to $\pi/2+ \delta$,
with $\delta\ne 0$ depending on the parameters.
Thus, unlike at the extremal points $\omega=0, \pi/2$, there are no
second-order transition lines.
Correspondingly, there are no values of the parameters for
which any of the pion masses vanish.
This is very different from the theory with
a $\tau_1$ twist, where we found a two-dimensional critical 
sheet~\cite{Horkel:2014nha} in ${\mhat_q,\ephat_q,\omega}$ space.

The absence of critical lines at nonextremal twist can be
understood in terms of symmetries. For $\omega=0$ and $\pi/2$,
the potential has a $\theta\to-\theta$ symmetry, and this $Z_2$
symmetry is broken by the condensate in the CP-violating
phase, leading to a massless pion at the transition.
For nonextremal twist, however, the potential
of Eq.~(\ref{eq:Varb}) has no such symmetry.
Lacking this symmetry, one expects, and finds, only first-order
transitions.

\section{Conclusions}
\label{sec:conc}

This work completes our study of how isospin breaking
impacts the phase structure of Wilson-like and twisted-mass fermions.
The main results are the phase diagrams presented in Figs.~\ref{fig:phaseold}
and \ref{fig:phasetwist}, together with the corresponding pion masses.
These results show how the combination of discretization errors and
nondegeneracy can bring unphysical phases closer to (or further away)
from the physical point.

The inclusion of electromagnetism into the analysis turns out to be
very straightforward, aside from the need to introduce independent
up and down critical masses. Electromagnetism has no impact on the
phase diagrams at leading order, because the condensates in the
CP-violating phases involve neutral pions. The only impact is to
uniformly increase the charged pion masses.

We have investigated within W\chpt\ 
the conditions used in Ref.~\cite{deDivitiis:2013xla} to
determine the two critical masses in the presence of electromagnetism.
We find that, unless one makes the electroquenched approximation,
the two conditions are in fact not independent.
To determine both critical masses one needs an additional
condition, and we have presented one possibility in Appendix~\ref{app:tune}.
Our condition requires simulating at nonzero (though small)
$\theta_{\rm QCD}$, and thus will be difficult to implement in practice,
but provides an existence proof that an alternative condition exists.

Our analysis has been carried out in infinite volume.
For the finite volumes used in lattice simulations one might
be concerned about significant finite-volume effects on the
electromagnetic contributions.
The impact on the results presented here, however, should be
minimal. The phase diagram will remain unaffected by electromagnetism,
while the shifts in critical masses are dominated by ultraviolet
momenta, themselves insensitive to the volume.
The only significant effect will be on electromagnetic mass shifts,
with $\cEM$ picking up an effective power-law volume
dependence~\cite{Hayakawa:2008an,Davoudi:2014qua,Borsanyi:2014jba,Basak:2014vca}.

\section*{Acknowledgments}
This work was supported in part by the United States Department of Energy 
grant DE-SC0011637.

\appendix
\section{Relating lattice masses to those in \chpt}
\label{app:renorm}

In this appendix we describe how bare lattice masses used in simulations with
Wilson-like fermions are related to the masses $m_u$ and $m_d$ appearing in 
\chpt\ (contained in the mass matrix $M$).
This discussion draws heavily from the results
of Ref.~\cite{Bhattacharya:2005rb}.
We do not consider the impact of electromagnetism here;
this is discussed in the subsequent appendix.

We must assume that the number of dynamical quarks in the 
underlying simulations is $N_f=3$ (up, down and strange)
or $N_f=4$ (adding charm). Working with up and down quarks 
alone turns out not to
be sufficient, but in any case this is not the physical theory.
We must also have that $a m_f \ll 1$ for all flavors $f$, 
so that an expansion in these quantities makes sense.
This condition is met by state-of-the-art simulations.
Note that this condition is much weaker
than the requirement that the quarks are light in the sense of \chpt,
which is $m_f \ll \Lambda_{\rm QCD}$.
In the main text,
we assume the latter condition holds only for up and down quarks.

Let $m_{0,f}$ be the bare dimensionless lattice mass for flavor $f$
(i.e. the mass appearing in the lattice action).
Because of the additive renormalization induced by explicit chiral symmetry
breaking, unrenormalized quark masses are given by
\begin{equation}
\widetilde m_f = \frac{m_{0,f}-m_{cr}}{a}
\,,
\end{equation}
where $m_{cr}$ is the (dimensionless) critical mass 
for the given number of dynamical flavors.
Methods to determine $m_{cr}$ are described below.
Then, as shown in Ref.~\cite{Bhattacharya:2005rb}, renormalized
masses are given by\footnote{%
The correction terms of ${\cal O}(a \tilde m^2)$
in (\ref{eq:massrenorm})
are subleading in our power-counting and will be dropped henceforth.}
\begin{equation}
m_f =  Z_m \left[\widetilde m_f + (r_m-1) \frac{\sum_f \widetilde m_f}{N_f}
+ {\cal O}(a \widetilde m^2)\right]
\,.
\label{eq:massrenorm}
\end{equation}
Here $Z_m$ is the renormalization constant for flavor nonsinglet
mass combinations such as $\epsilon_q=(m_u\!-\!m_d)/2$,
while $Z_m r_m$ is the corresponding constant for the average quark mass.
$r_m-1$ is a finite constant, 
arising first at ${\cal O}(g^4)$ in perturbation theory.
By implementing continuum
Ward-Takahashi identities, one can determine $r_m$ nonperturbatively
for $N_f=3$ and $4$, although not for $N_f=2$~\cite{Bhattacharya:2005rb}.
This is the reason for the restriction on $N_f$ noted above.
We assume here that $r_m$ has been calculated in this way.

Equation~(\ref{eq:massrenorm}) shows that the renormalized mass $m_f$
does not vanish when $\tilde m_f=0$ if other flavors are massive.
Specifically, for the up and down quarks we have
\begin{align}
m_u+m_d &= Z_m \frac{1\!+\!r_m}2 (\widetilde m_u+\widetilde m_d)
\nonumber\\
&\ \ + Z_m \frac{r_m\!-\!1}2 (\widetilde m_s+\widetilde m_{ch})\,,
\label{eq:masssumrenorm}
\\
m_u-m_d &= Z_m (\tilde m_u-\tilde m_d)\,.
\label{eq:massdiffrenorm}
\end{align}
(Here we we have chosen $N_f=4$ for definiteness; the result for $N_f=3$
is similar.)
Thus the two-flavor massless point receives an overall additive
shift due to the strange and charm quarks,
and we also see explicitly the difference between singlet and nonsinglet
renormalizations.

These results imply that, in terms of unrenormalized masses,
the phase diagrams of Fig.~\ref{fig:phaseold}
would be translated in the vertical
direction (due to the additive mass shift) and stretched by {\em different}
factors in the vertical and horizontal directions.
The respective stretch factors are $B_0 Z_m(1+r_m)/2$ and $B_0 Z_m$.
If, however, $r_m$ is known, then the two stretch factors can
be made equal by applying a finite renormalization to remove the $(1+r_m)/2$
factor. Knowledge of $Z_m$ is, however, not useful,
since it always appears multiplied by the unknown LEC $B_0$.

We would like to be able to remove
the additive mass shift in Eq.~(\ref{eq:masssumrenorm}).
To do so we consider how the critical mass $m_{cr}$ is determined.
The expressions above assume that it has been obtained
by doing simulations with $N_f$ {\em degenerate} quarks of mass $m$,
and equating $m_{cr}$ to the value of $m$ at which
the ``PCAC mass'' vanishes. This is equivalent to imposing
\begin{equation}
\langle \pi^+| \partial_\mu (\bar u \gamma_\mu \gamma_5 d) 
| 0\rangle\Big|_{m=m_{cr}}  = 0
\,.
\end{equation}
If, instead, one imposes this condition by varying $m=m_u=m_d$,
with $m_s$ and $m_{ch}$ held fixed at their physical values,
then the $m_{cr}$ so obtained automatically includes the shift
due to loops of strange and charm quarks.
This is because one is enforcing a consequence of
chiral symmetry in the two-flavor subsector.
With this new choice of $m_{cr}$, and with the adjustment
of stretch factors described above, the phase
diagrams of Fig.~\ref{fig:phaseold} apply directly
for lattice masses $\tilde m_f$. 

This new choice of $m_{cr}$ has a second advantage: it removes
an additional shift of $O(a)$ in the relation between bare
quark masses and the masses appearing in \chpt. 
As explained in Ref.~\cite{Sharpe:1998xm},
this shift is caused by the ${\cal O}(a)$
clover term in the Symanzik effective action
(and is thus absent for nonperturbatively improved Wilson fermions).
In the main text it is assumed that this shift
has been removed.

Since we include ${\cal O}(a^2)$ terms in the main text,
we must determine how they impact the considerations above.
There is no further shift in the quark masses at this order---this
next occurs at ${\cal O}(a^3)$~\cite{Sharpe:2005rq}.
However, as illustrated by Fig.~\ref{fig:phaseold},
the ${\cal O}(a^2)$ terms do impact the phase diagram.
This means that, in general, one cannot use the vanishing
of the PCAC mass to determine $m_{cr}$ with untwisted Wilson fermions. 
For example, if one is in the first-order scenario 
[Fig.~\ref{fig:phaseold}(b) along the $\hat m_q$ axis],
then the PCAC mass simply does not vanish for any $\hat m_q$.
Instead, one must introduce a twisted component to the mass,
$\mu\sim {\cal O}(a)$, and then enforce the vanishing of the
PCAC mass (in the so-called ``twisted basis''). Extrapolating
the result linearly to $\mu=0$ yields a result for $m_{cr}$
that has errors of ${\cal O}(a^3)$, which is sufficiently accurate
for our analysis. 
For a detailed discussion of this point see Ref.~\cite{Sharpe:2005rq}.

In summary, by determining $r_m$ from Ward identities,
and the critical mass from the PCAC mass condition with
twisted-mass quarks, one can obtain lattice quark masses which
are proportional to those appearing in \chpt\ at the
order we work. Specifically, we find
\begin{equation}
\frac{\mqhat}{B_0 Z_m} = \frac{1\!+\!r_m}2 (\tilde m_u\!+\!\tilde m_d)\ 
{\rm and}\ \
\frac{\epqhat}{B_0 Z_m} = (\tilde m_u\!-\!\tilde m_d)\,,
\label{eq:untwistedZ}
\end{equation}
where $\mqhat$ and $\epqhat$ are the quantities appearing in the chiral
potential of Eq.~(\ref{eq:Vchi2}).

This analysis can be straightforwardly extended to arbitrary twist.
We begin with maximal twist, for which the mass matrix in \chpt\ is given
by Eq.~(\ref{eq:chimaxtwist}), and the relevant bare masses are $\mu_0$ and $\eta_0$
of Eq.~(\ref{eq:twistedlatticeaction}).
In this case there is no additive renormalization, but the presence
of different renormalization factors for singlet and nonsinglet
masses remains. 
Using the results of Ref.~\cite{Bhattacharya:2005rb}, we find\footnote{%
Specifically, we have used $Z_m=1/Z_S$ and $r_m=1/r_S$.}
\begin{equation}
\frac{\mqhat}{B_0 Z_m} = \frac{Z_S}{Z_P}\frac{1\!+\!r_P}{r_P}\mu_0\  \  
{\rm and}\ \
\frac{\epqhat}{B_0 Z_m} =\frac{Z_S}{Z_P} \eta_0 \,.  
\label{eq:twistedZ}
\end{equation}
Here $Z_S/Z_P$ and $r_P$ are finite constants,
both of which can be determined from
Ward identities for $N_f=3$ and $4$, 
but not for $N_f=2$~\cite{Bhattacharya:2005rb}.
Like $r_m$, $r_P$ begins at ${\cal O}(g^4)$ in perturbation theory. 

At arbitrary twist one has four bare masses, and they are related
to the corresponding four renormalized masses using the same renormalization
factors as given in Eqs.~(\ref{eq:untwistedZ}) and (\ref{eq:twistedZ}).

Finally, we stress that the analysis presented here does not include
electromagnetic effects. The dominant such effect is that
the critical mass $m_{cr}$ has to be chosen differently for the
up and down quarks, and is discussed in the following appendix. 
A subdominant, but still important, effect is that
the renormalization factors now depend not only on $\alpha_S$ but also on
$\alpEM$. The latter dependence can presumably be adequately captured
using perturbation theory. 
The formulae given above still hold
if one uses the new critical masses and renormalization factors. 

\section{Determining the critical masses in the presence of electromagnetism}
\label{app:tune}

The analysis of the previous appendix must be extended when electromagnetism
is included, due to the presence of charge-dependent self energy 
corrections proportional to $\alpEM/a$.
This implies that the critical masses for up and down quarks differ,
and we label them $m_{cr,u}$ and $m_{cr,d}$, respectively.
In Ref.~\cite{deDivitiis:2013xla} two methods for a nonperturbative
determination of these critical masses are proposed.
One of these (the method used in practice in Ref.~\cite{deDivitiis:2013xla})
involves only up and down quarks, and thus
can be implemented, and therefore checked, within SU(2) W\chpt. 
We do so in this appendix, finding that
{\em the method does not fix both critical masses}, but rather
constrains them to lie in a one-dimensional subspace of the
$m_{cr,u}$---$m_{cr,d}$ plane. We then provide an additional condition that
does completely determine $m_{cr,u}$ and $m_{cr,d}$.

The tuning conditions require the use of twisted-mass quarks,
although the resulting values of $m_{cr,u}$ and $m_{cr,d}$
apply for both Wilson and twisted-mass quarks.
Thus the lattice quark Lagrangian is given by 
Eq.~(\ref{eq:twistedlatticeaction}).
We can write the mass matrix in two useful forms
\begin{multline}
 m_{0} + \tau_3 \epsilon_{0}
+ i \gamma_5 \tau_3 \mu_0  + i \gamma_5 \eta_{0}
\\
=
\begin{pmatrix}
m_{0,u} + i \gamma_5 \mu_{0,u} & 0\\
0 & m_{0,d} - i \gamma_5 \mu_{0,d} \\
\end{pmatrix}\,.
\label{eq:newmassmatrix}
\end{multline}
The tuning proceeds by first choosing
bare twisted masses $\mu_{0,u}$ and $\mu_{0,d}$ such that, 
when multiplicatively renormalized as described in the previous appendix, 
they give rise, respectively, to the
desired physical up and down quark masses.\footnote{%
In fact, the tuning can be done using any values of the twisted
masses which respect our power counting. The critical masses do
not depend on the twisted masses at the order we work.}
The negative sign multiplying $\mu_{0,d}$ is chosen to correspond
to a $\tau_3$ twist.
The second step is to tune the untwisted masses $m_{0,u}$ and $m_{0,d}$ 
to their critical values such that the (additively) renormalized 
untwisted masses vanish.

The method of determining $m_{cr}$ used in the previous section
is no longer useful---the vanishing of the PCAC mass is
a condition based on the recovery of the chiral SU(2) group,
but this group is explicitly broken by electromagnetism.
The workaround proposed in Ref.~\cite{deDivitiis:2013xla} is
to add to the sea quarks (labeled $u_S$ and $d_S$) a pair of
valence quarks, $u_V$ and $d_V$, each of which has the same
charge and untwisted mass as the corresponding sea quark, 
but has opposite twisted mass.\footnote{%
This description is equivalent to that 
of Ref.~\cite{deDivitiis:2013xla}, but differs technically in two ways.
First, we find that one need only introduce two valence quarks to describe
the method, 
rather than the four used in Ref.~\cite{deDivitiis:2013xla}.
This does not impact the method itself, only its description.
Second, we work in the twisted basis, rather than the physical
basis used in Ref.~\cite{deDivitiis:2013xla}.}
Thus $(u_S,u_V)$ and $(d_V,d_S)$ each form a twisted pair.
The key point is that, within each pair, the
${\cal O}(\alpEM/a)$ shift in the untwisted mass is common. Therefore
it is plausible that one can determine the critical mass for each pair
by enforcing the recovery of the corresponding valence-sea chiral SU(2).
Specifically, $m_{cr,u}$ is determined by
\begin{equation}
\langle \pi^u_{SV}| \partial_\mu (\bar u_S \gamma_\mu \gamma_5 u_V) 
| 0\rangle\Big|_{m_{0,u}=m_{cr,u}}  = 0 \,,
\label{eq:utune}
\end{equation}
while $m_{cr,d}$ is determined by 
the analogous condition with $u\to d$:
\begin{equation}
\langle \pi^d_{SV}| \partial_\mu (\bar d_S \gamma_\mu \gamma_5 d_V) 
| 0\rangle\Big|_{m_{0,d}=m_{cr,d}}  = 0
\,.
\label{eq:dtune}
\end{equation}
Here $\pi^u_{SV}$ and $\pi^d_{SV}$ are sea-valence pions composed,
respectively, of up and down quarks.

When using a partially quenched theory, one also needs to
add ghost fields, $\tilde u_V$ and $\tilde d_V$, to cancel
the valence quark determinants.\footnote{%
For reviews of partially
quenched theories and the corresponding \chpt, see 
Refs.~\cite{Sharpe:2006pu,Golterman:2009kw}.}
Thus the full softly-broken chiral symmetry is the graded group
$SU(4|2)_L \times SU(4|2)_R$.
This raises the question of whether complications arising
from partial quenching, or from discretization effects,
can lead to corrections to the
tuning criteria of Eqs.~(\ref{eq:utune}) and (\ref{eq:dtune}).
This is one of
the issues we address here by mapping these conditions into \chpt.

We begin by mapping the mass matrix in the unquenched sector
into \chpt. The four parameters of Eq.~(\ref{eq:newmassmatrix})
map into
\begin{align}
\chi
&=
\begin{pmatrix}
\hat m_u e^{i\omega_u} & 0\\
0 & \hat m_d e^{-i\omega_d} \\
\end{pmatrix} 
\\
&=
\begin{pmatrix}
(\mhat_q+\ephat_q)e^{i(\omega+\varphi)} & 0\\
0 & (\mhat_q-\ephat_q)e^{i(-\omega+\varphi)} \\
\end{pmatrix}\,.
\label{eq:chiincludingphi}
\end{align}
The choice of sign for $\omega_d$ is such that it is positive
with a $\tau_3$ twist.
$\chi$ contains the additional parameter $\varphi$ compared to
the mass matrix analyzed in the main text, Eq.~(\ref{eq:chitwisted}).
$\varphi$ is a measure of the difference between 
up and down twist angles,
\begin{equation}
\omega_u=\omega+\varphi\,, \qquad
\omega_d=\omega-\varphi\,.
\end{equation}
As discussed in Sec.~\ref{sec:twistchiral}, such a difference corresponds to
the introduction of a nonzero $\thetaQCD$---the
explicit relation is $\varphi=\thetaQCD/2$.

We note that the relations between the bare masses 
of Eq.~(\ref{eq:newmassmatrix}) and the parameters of $\chi$ in
Eq.~(\ref{eq:chiincludingphi})---which can be worked out along the lines of
the previous appendix---are not needed here.
All we need to know is that, if $m_{0,u}=m_{cr,u}$ and
$m_{0,d}=m_{cr,d}$, then both up and down masses are fully
twisted. Thus the twist angles in $\chi$ are
$\omega_u=\omega_d=\pi/2$, implying maximal twist
with no $\theta_{\rm QCD}$ term: $\omega=\pi/2$ and $\varphi=0$.
Reaching this point in parameter space is the aim of tuning.

When considering the PQ extension of this theory, we will focus
mainly on the quark sector, since the ghosts do not play a
significant role. Collecting the four quark fields in the order
\begin{equation}
\psi_{\PQ}^\top = (u_S, u_V, d_V, d_S)\,,
\end{equation}
the extended quark mass matrix is
\begin{equation}
\chi_{\PQ}=\begin{pmatrix}
(\mhat_q+\ephat_q)e^{i\omega_u \tau_3} & 0\\
0 & (\mhat_q-\ephat_q)e^{i\omega_d \tau_3} \\
\end{pmatrix}\,.
\label{eq:chiPQ}
\end{equation}
The factors of $\tau_3$ arise because, by construction,
valence quarks have opposite twisted masses to the corresponding sea quarks.
We stress that the ${\cal O}(\alpEM/a)$ shifts are incorporated
into the parameters $\mhat_q$ and $\ephat_q$, along with the
usual ${\cal O}(1/a)$ shifts. We can also include the
${\cal O}(a)$ shifts in the same fashion. 

To implement the conditions (\ref{eq:utune}) and (\ref{eq:dtune})
in the PQ theory, we need the extension of $\Sigma$ to this theory.
This is a $6\times 6$ matrix transforming in the usual way
under $SU(4|2)_L \times SU(4|2)_R$.
In fact, as we only need matrix elements for states composed of
quarks, and since we know from Ref.~\cite{Bar:2010ix}
that there are no quark-ghost condensates, 
we can focus on the $4\times 4$ quark sub-block,
which we call $\Sigma_{\PQ}$.
We now argue that the expectation value of $\Sigma_{\PQ}$
has the form
\begin{equation}
\langle\Sigma_{\PQ}\rangle 
= {\rm diag}(e^{i\theta}, e^{-i\theta},e^{i\theta},e^{-i\theta})
\,.
\label{eq:SigmaPQ}
\end{equation}
This is based on the following results.
First, the unquenched $2\times 2$ block of $\Sigma_{\PQ}$
(i.e. that involving the first and last rows and columns) 
is just the unquenched $\Sigma$ field.
This is unaffected by partial quenching~\cite{Sharpe:2000bc,Sharpe:2001fh},
and its expectation value is given by an unquenched \chpt\ 
calculation. This calculation must 
include not only nondegeneracy, electromagnetism and twist,
but also nonvanishing $\thetaQCD$.
To our knowledge such an analysis has not previously been
done, so we carry it out in a companion paper~\cite{chptforthetaqcd}.
The result is that the unquenched condensate $\langle\Sigma\rangle$
only rotates in the $\tau_3$ direction---there are no
off-diagonal condensates such as $\langle\bar u_S d_S\rangle$.
This fixes the first and last entries in Eq.~(\ref{eq:SigmaPQ})
to have opposite phase angles.

This is an important result for the following, so we emphasize
its key features. Although $\thetaQCD\ne 0$ leads to an overall
phase in the mass matrix [$e^{i\varphi}$ in Eq.~(\ref{eq:chiincludingphi})],
its effect on the condensate $\langle\Sigma\rangle$ is qualitatively
similar to that of a twist $\omega$, despite the fact that the
latter leads to opposite phases on $u$ and $d$ quarks. This happens
because $\Sigma$ is constrained to lie in $SU(2)$, and so has
no way to break parity other than rotating in the $\tau_3$
direction. An overall phase rotation would take it out of $SU(2)$
into the $U(2)$ manifold.

The second result needed to obtain Eq.~(\ref{eq:SigmaPQ}) is
the existence of relations between valence and sea-quark condensates.
In particular, one can show that
\begin{equation}
\langle \bar u_V u_V\rangle=\langle \bar u_S u_S \rangle
\ \ {\rm and}\ \
\langle \bar u_V \gamma_5 u_V\rangle=-\langle \bar u_S \gamma_5 u_S \rangle
\,,
\label{eq:condVvsS}
\end{equation}
to all orders in the hopping parameter expansion.
The minus sign in the second relation follows from the
opposite twisted mass of sea and valence quarks.
The result (\ref{eq:condVvsS}) 
holds on each configuration and thus also for the ensemble average,
even though the measure is complex for $\thetaQCD\ne0$.
Since the additive and multiplicative renormalizations of these
condensates are the same for valence and sea quarks,
the result (\ref{eq:condVvsS}) 
implies that valence and sea up-quark condensates have opposite ``twists'',
$e^{\pm i\theta}$. 
The same argument applies to the down-quark condensates,
and taken together these arguments determine the form of the
second and third diagonal elements in Eq.~(\ref{eq:SigmaPQ}).

The final result needed to obtain the form (\ref{eq:SigmaPQ})
is the vanishing of off-diagonal condensates involving one or more valence
quarks, e.g. $\langle \bar u_V d_V\rangle$ and $\langle \bar u_V d_S\rangle$.
These differ from the diagonal condensates in that there
is no mass term in the quark-level Lagrangian that can serve as a source
for such condensates. Thus to determine whether they are nonzero
one must add a source,
e.g. $\Delta \,\bar d_V u_V$, calculate the resulting condensate,
send the volume to infinity, and finally send the parameter $\Delta\to 0$.
This analysis has been carried out in Appendix A of
Ref.~\cite{Hansen:2011kk} in a theory with twisted-mass quarks,
although, unlike our situation, the
quarks were degenerate and $\thetaQCD=0$.
The general lessons from Ref.~\cite{Hansen:2011kk} are 
(i) that to obtain a nonvanishing condensate one needs
a source of infrared divergence to cancel the overall factor of $\Delta$,
and (ii) that nonvanishing twisted masses cut off such divergences.
These lessons apply also for all the off-diagonal condensates that we 
consider here.
However, the argument as given in Ref.~\cite{Hansen:2011kk} assumes that
the measure is real and positive, which does not hold here.
Nevertheless, since we are tuning to $\thetaQCD=0$, we expect the
impact of having a complex measure to be small.
Furthermore, we know from Ref.~\cite{chptforthetaqcd}
that the corresponding sea quark condensates, 
e.g. $\langle \bar u_S d_S\rangle$ and
$\langle\bar u_S \gamma_5 d_S\rangle$, 
vanish even when $\thetaQCD\ne0$.
These condensates differ from those containing valence quarks
only by changing the signs of some of the twisted masses.
Since it is the presence of these masses, and not their detailed
properties, that leads to the vanishing of the condensate,
we expect the result holds for all off-diagonal condensates.

With the form (\ref{eq:SigmaPQ}) in hand, we can now apply 
the tuning conditions (\ref{eq:utune}) and (\ref{eq:dtune}) in \chpt.
We do so by generalizing the analysis of Ref.~\cite{Sharpe:2004ny},
where the twist angle for unquenched twisted-mass fermions
was determined in \chpt\ by applying a PCAC-like condition.
The required extension is from the $SU(2)$ sea-quark sector alone to
the full valence-sea $SU(4)$ symmetry.
Much of the analysis carries over with minimal changes from
Ref.~\cite{Sharpe:2004ny}, so we only sketch the calculation.

The first step is to obtain the pion fields that couple
to external particles in the tuning conditions.
Following Ref.~\cite{Sharpe:2004ny}, we obtain these by
expanding the chiral field about its vacuum value as
\begin{align}
\Sigma_{\PQ} &= 
\xi_\PQ\, e^{2 i\Pi/f} \, \xi_\PQ\,,
\\
\Pi &= \sum_{a=1}^{15} \pi_a \lambda_a\,,
\\
\xi_\PQ &= 
\sqrt{\langle\Sigma_{\PQ}\rangle} =
{\rm diag}(e^{i\theta/2}, e^{-i\theta/2},e^{i\theta/2},e^{-i\theta/2}) 
\,.
\label{eq:sigmaPQpion}
\end{align}
Here $\lambda_a$ are the generators of SU(4), with $\pi_a$ 
the corresponding pion fields. These are the pions in the PQ
theory that are composed of quarks alone, with no ghost component.\footnote{%
A similar form to Eq.~(\ref{eq:sigmaPQpion})
holds for the full $6\times6$ PQ chiral field,
but we can focus on the $SU(4)$ block, since the pions we
leave out in this way are those containing one or more ghost fields.}
The pions needed for tuning,
$\pi^u_{SV}$ and $\pi^d_{SV}$, 
are contained in the upper and lower diagonal $2\times2$ blocks of
$\Pi$, respectively.

The next step is to determine the form, in \chpt,
of the axial currents appearing in the tuning conditions.
These can be obtained by introducing sources into derivatives
using standard methodology. Since, by definition, our chiral potential
does not contain derivatives, at \LOP\ only the LO kinetic term
[shown in Eq.~(\ref{eq:Lchi})]
enters into the determination of the currents.
We do not display the form of the currents, however,
as the calculation needed for each of the tuning conditions
is {\em exactly} the same as that carried out in Ref.~\cite{Sharpe:2004ny}.
This is because each tuning condition involves
a separate, nonoverlapping $SU(2)$ subgroup of $SU(4)$
(upper-left or lower-right $2\times 2$ block), 
and because the condensate
(\ref{eq:SigmaPQ}) does not connect these subgroups.
We simply quote the results of the calculation:
\begin{align}
\langle \pi^u_{SV}| \partial_\mu (\bar u_S \gamma_\mu \gamma_5 u_V) 
| 0\rangle  & \propto \cos\theta \,,
\label{eq:utune2}
\\
\langle \pi^d_{SV}| \partial_\mu (\bar d_S \gamma_\mu \gamma_5 d_V) 
| 0\rangle & \propto \cos\theta \,.
\label{eq:dtune2}
\end{align}
Thus enforcing either (\ref{eq:utune}) or (\ref{eq:dtune}) has
the effect of setting $\theta=\pm \pi/2$ and the condensate to
\begin{equation}
\langle\Sigma_\PQ\rangle=\pm {\rm diag}(i,-i,i,-i)\,,
\end{equation}
For our choices of signs of the twisted masses $\mu_{0,u}$
and $\mu_{0,d}$ in Eq.~(\ref{eq:newmassmatrix}),
the $\pm$ signs are in fact plusses, i.e. $\theta=\pi/2$.

A surprising aspect of this result is that the two tuning conditions 
are not independent:
if one enforces, say, Eq.~(\ref{eq:utune}) then Eq.~(\ref{eq:dtune})
will be automatically satisfied.
This dependence arises because changing
$m_u$ in turn changes $\varphi$ and $\omega$ and 
this impacts the $d$ condensate through the quark determinant.
One might, therefore, wonder how the two tuning conditions 
have been successfully applied in Ref.~\cite{deDivitiis:2013xla}.
To understand this, we note that this work makes two approximations.
First, isospin-breaking effects are evaluated only through linear
order in an expansion in $m_u\!-\!m_d$ and $\alpEM$.
Second, insertions of $m_u\!-\!m_d$ or photons on sea-quark loops
are dropped (the ``electroquenched approximation'').
The latter approximation has the effect of
disconnecting the two tuning conditions---all quark loops 
in both conditions are evaluated with uncharged, degenerate sea-quarks,
so the $u$-quark condensate cannot be impacted by changes in $m_d$
and vice versa. Since \chpt\ predicts that there is a tight correlation
between the condensates, it appears to us that the
electroquenched approximation is theoretically problematic.
However, from a purely numerical viewpoint, the dropped contributions
may well lead only to small corrections.

\bigskip
The lack of independence implies that the tuning conditions 
cannot determine both $m_{u,c}$ and $m_{d,c}$---only one
constraint on these two critical masses is obtained.
In terms of the parameters of mass matrix (\ref{eq:chiincludingphi}),
the conditions determine only a relation between $\omega$ and $\varphi$.
Thus, after enforcing either (\ref{eq:utune}) or (\ref{eq:dtune})
the theory is known to lie along a line in the 
$\omega$---$\varphi$ plane.
In terms of the bare masses, the theory lies along a line
in the $m_{0,u}$---$m_{0,d}$ plane (with, recall,
$\mu_{0,u}$ and $\mu_{0,d}$ fixed at the values leading to
physical quark masses when $m_{0,u}=m_{0,d}=0$).
We do know that this one-dimensional subspace includes 
the point we are trying to tune to, namely that with
$(\omega,\varphi)=(\pi/2,0)$. 
This follows from the analysis of Sec.~\ref{subsec:maxtwist}.
At maximal twist with $\varphi=0$, the twist in the condensate
is also maximal, i.e. $\theta=\pi/2$.
The only caveat is that the values of the twisted masses
must be such that one lies in the continuum-like phase, rather
than the CP-violating phase (see Fig.~\ref{fig:phasetwist}).

To complete the tuning we need an additional condition that
forces us to the desired point along the allowed line.
At first blush one might expect that it would be simple
to find an additional tuning condition, since theories with
$\thetaQCD\ne0$ have explicit parity violation.
This is in contrast to the parity violation induced by a nonzero
twist $\omega$ which, in the continuum limit,
can be removed by a chiral rotation.
This suggests that one should look for quantities
that vanish when parity is a good symmetry. 
The flaw in this approach is that parity is broken by $\omega\ne 0$
away from the continuum limit---the chiral rotation required to
obtain the parity-symmetric form is not a symmetry on the lattice.
Thus the distinction between $\varphi\ne0$ and $\omega\ne 0$ no longer holds.

The only choice that we have found for a second condition involves
using the pion masses. Specifically, we find that,
along the line picked out by setting $\theta=\pi/2$, 
the masses of both charged and neutral pions are minimized 
when $\varphi=0$. This assumes only
that we are in the continuum-like phase for the physical values
of $\mu_{0,u}$ and $\mu_{0,d}$.

The details of the calculation are presented in 
Ref.~\cite{chptforthetaqcd}.
Working at \LOP, we find that the constraint $\theta=\pi/2$
forces the quark masses to lie on the line
\begin{equation}
\frac{\hat m_d}{\hat m_u}
=
- \left( \frac{1- c_{\ell} (\hat \mu_u - \hat \mu_d)}
              {1+ c_{\ell} (\hat \mu_u - \hat \mu_d)} \right)
\,.
\end{equation}
As noted above, this line passes through the desired
point $m_u=m_d=0$. The slope is $-1$ when $c_\ell=0$, 
and increases in magnitude as $c_\ell$ increases
(assuming the physical situation $\hat \mu_u < \hat \mu_d$).
There is no singularity when the slope reaches infinity---this
simply means that the constraint line is the $m_u=0$ axis.
For larger $c_\ell$ the slope is positive.
It decreases with increasing $c_\ell$, 
though it always remains greater than unity.
The pion masses along the constraint line are 
\begin{align}
m_{\pi^0}^2&= \frac{\hat \mu_u\!+\!\hat \mu_d}2
-2 c_\ell \left(\frac{\hat \mu_u\!-\!\hat \mu_d}2\right)^2
\nonumber\\
&\qquad \ \ + 2 c_\ell \left(\frac{\hat m_u\!-\!\hat m_d}2\right)^2 -2 w'\,,
\label{eq:mpi0constrained}
\\
m_{\pi^\pm}^2&= \frac{\hat \mu_u\!+\!\hat \mu_d}2
+ 2 c_\ell \left(\frac{\hat m_u\!-\!\hat m_d}2\right)^2
+ 2 \cEM\,.
\label{eq:mpipmconstrained}
\end{align}
Thus we see that both masses are minimized along the constraint
line when $m_u=m_d=0$. 
If one were to implement this tuning condition in practice,
then one would apply it for the charged pion masses,
since these have no quark-disconnected contractions.

This analysis breaks down when $c_\ell$ gets too large,
because the theory with $m_u=m_d=0$ then lies in the the CP-violating phase. 
This can be seen from the result (\ref{eq:mpi0constrained})---for large
enough $c_\ell$ the squared neutral pion mass becomes negative.
This happens sooner for the first-order scenario, $w'>0$.

We close this section by commenting on the impact of higher-order
terms in \chpt. Because of such terms, even if one perfectly implements
our two tuning conditions---namely either Eq.~(\ref{eq:utune})
or (\ref{eq:dtune}) and minimizing the pion masses---one will not
have precisely tuned to $m_u=m_d=0$.
This can be seen, for example, from the analysis of 
Ref.~\cite{Sharpe:2004ny}, where terms of ${\cal O}(a p^2, a m)$ lead to
maximal twist occurring at untwisted masses of ${\cal O}(a \mu)$,
with $\mu$ the twisted mass, rather than zero. 
Shifts of this size occur also in the presence of isospin
breaking, although the detailed form of the corrections will differ.
Within our power-counting, however, $\mu \sim a^2$ so that the
shifts in the untwisted masses are $\sim {\cal O}(a^3)$, 
beyond the order that we consider.

\bibliographystyle{apsrev4-1} 
\bibliography{ref}

\end{document}